\documentclass[a4paper,11pt]{article}
\pdfoutput=1 

\usepackage{jcappub} 
\usepackage{indentfirst}
\usepackage{appendix}
\usepackage{multirow}
\usepackage{graphicx}
\usepackage{subfigure}

\newcommand{\be}{\begin{equation}}
\newcommand{\ee}{\end{equation}}
\newcommand{\bq}{\begin{eqnarray}}
\newcommand{\eq}{\end{eqnarray}}
\newcommand{\n}{\nonumber}

\def\({\left(}
\def\){\right)}
\def\[{\left[}
\def\]{\right]}

\title{A forecast of using fast radio burst observations to constrain holographic dark energy}

\author[a]{Xing-Wei Qiu,}
\author[a]{Ze-Wei Zhao,}
\author[a]{Ling-Feng Wang,}
\author[a]{Jing-Fei Zhang}
\author[a,b,c,1]{and Xin Zhang\note{Corresponding author.}}
\affiliation[a]{Department of Physics, College of Sciences, Northeastern
University, Shenyang 110819, China}
\affiliation[b]{MOE Key Laboratory of Data Analytics and Optimization for Smart Industry, Northeastern University, Shenyang 110819, China}
\affiliation[c]{Frontiers Science Center for Industrial Intelligence and Systems Optimization, Northeastern University, Shenyang 110819, China}

\emailAdd{qiuxingwei@stumail.neu.edu.cn,
zhaozw@stumail.neu.edu.cn,
lingfengwang@stumail.neu.edu.cn,
jfzhang@mail.neu.edu.cn, zhangxin@mail.neu.edu.cn}
\abstract{
Recently, about five hundred fast radio bursts (FRBs) detected by CHIME/FRB Project have been reported. The vast amounts of data would make FRBs a promising low-redshift cosmological probe in the forthcoming years, and thus the issue of how many FRBs are needed for precise cosmological parameter estimation in different dark energy models should be detailedly investigated. Different from the usually considered $w(z)$-parameterized models in the literature, in this work we investigate the holographic dark energy (HDE) model and the Ricci dark energy (RDE) model, which originate from the holographic principle of quantum gravity, using the simulated localized FRB data as a cosmological probe for the first time. We show that the Hubble constant $H_0$ can be constrained to about 2\% precision in the HDE model with the Macquart relation of FRB by using 10000 accurately-localized FRBs combined with the current CMB data, which is similar to the precision of the SH0ES value. Using 10000 localized FRBs combined with the CMB data can achieve about 6\% constraint on the dark-energy parameter $c$ in the HDE model, which is tighter than the current BAO data combined with CMB. We also study the combination of the FRB data and another low-redshift cosmological probe, i.e. gravitational wave (GW) standard siren data, with the purpose of measuring cosmological parameters independent of CMB. Although the parameter degeneracies inherent in FRB and in GW are rather different, we find that more than 10000 FRBs are demanded to effectively improve the constraints in the holographic dark energy models.}

\begin{document}

\maketitle

\section{Introduction} \label{sec:intro}
In 1998, two independent studies of distant supernovae discovered the accelerated expansion of the universe \cite{Riess:1998cb,Perlmutter:1998np}, which is one of the most surprising astronomical discoveries in history. One can explain the accelerated expansion by introducing a component with negative pressure, usually called dark energy, and study its properties by cosmological observations. The cosmic microwave background (CMB) data measured by the \emph{Planck} satellite \cite{Aghanim:2018eyx} provided precise constraints on the cosmological parameters in the $\Lambda$ cold dark matter ($\Lambda$CDM) model, which is usually regarded as the standard model of cosmology \cite{Bahcall:1999xn} with the origin of dark energy being explained as the cosmological constant $\Lambda$.

The $\Lambda$CDM model fits the current cosmological observations quite well, but it suffers from two theoretical puzzles: the fine-tuning problem and the coincidence problem \cite{Weinberg:1988cp}. Some dynamical dark energy models may relieve these problems \cite{Joyce:2014kja,Zhao:2018fjj,Feng:2019mym,Li:2018ydj}.
However, the extra dark-energy parameters are hard to be precisely constrained by the CMB data alone due to the strong parameter degeneracies \cite{Akrami:2018vks}.
The baryon acoustic oscillation (BAO) measurements from galaxy redshift surveys \cite{Beutler:2011hx,Ross:2014qpa,Alam:2016hwk} as a representative low-redshift cosmological probe are usually combined with the CMB data to break the cosmological parameter degeneracies. In the future, some other low-redshift cosmological probes are expected to yield large amounts of data, such as fast radio bursts (FRBs) and gravitational waves (GWs). Therefore, what precision to cosmological parameters these two non-optical probes could measure and how many events are needed for precise cosmological parameter estimation in different dark energy models are important questions at present.

FRBs are extremely bright, short-duration radio
signals. One of the important characteristics of FRBs is the high dispersion measure (DM), which contains valuable information on the cosmological distance they have traveled. In 2007, the first FRB event, FRB010724, was observed by the 64-m Parkes Radio Telescope in Australia \cite{Lorimer:2007qn}. Although FRB010724 has a high Galactic latitude, it was previously thought to be caused by artificial interference due to its low signal-to-noise ratio (SNR). The second FRB event, FRB010621, was reported in ref.~\cite{Keane:2012yh} in 2012. Later in 2013, Thornton et al. reported four new FRB samples \cite{Thornton:2013iua}, which makes the study of FRBs become an important new direction in radio astronomy.
Recently, the Canadian Hydrogen Intensity Mapping Experiment (CHIME)/FRB Project has released its first catalog of 535 FRBs, including 61 bursts from 18 previously reported repeating sources \cite{Amiri:2021tmj}. Till now, 19 FRBs' host galaxies and cosmological redshifts have been identified \cite{Spitler:2016dmz,Chatterjee:2017dqg,Tendulkar:2017vuq,Kokubo:2017kkg,Bassa:2017tke,Prochaska231,Ravi:2019alc,Bannister:2019iju,CHIMEFRB:2020bcn,chittidi2020dissecting,Bhandari:2020oyb,mannings2020high,Marcote:2020ljw,Macquart:2020lln,James:2021oep,Bhardwaj:2021xaa,Law:2020cnm,heintz2020host,simha2020disentangling}. These abundant data tremendously exceed the amount of current data of GWs with electromagnetic (EM) counterparts and further prove that FRBs may become a promising cosmological probe in the future.

The data of localized FRBs could be used to constrain cosmological parameters with the
Macquart relation, which provide a relationship between $\rm DM_{\rm IGM}$ and $z$ \cite{Macquart:2020lln}. There have been a series of works using FRBs as a cosmological probe to study the expansion history of the universe, such as estimating the cosmological parameters \cite{Deng:2013aga} and conducting cosmography \cite{Gao:2014iva} by using FRBs and Gamma-Ray Bursts association, using FRBs to measure Hubble parameter $H(z)$ \cite{Wu:2020jmx} and the Hubble constant \cite{Hagstotz:2021jzu,Wu:2021jyk}, using the FRB data combined with the BAO data or the type Ia supernovae (SN) data to break the cosmological parameter degeneracies \cite{Zhou:2014yta,Jaroszynski:2018vgh},
using FRBs to probe diffuse gas \cite{Walters:2019cie}, using FRBs foreground mapping to constrain the cosmic baryon distribution \cite{Lee:2021ppm}, using FRBs to detect helium reionization \cite{Linder:2020aru},
using FRB dispersion measures as distance measures \cite{Kumar:2019qhc}, probing compact dark matter \cite{Munoz:2016tmg,Wang:2018ydd}, testing Einstein's weak equivalence principle \cite{Wei:2015hwd,Yu:2018slt,Xing:2019geq}, and many other works \cite{Yang:2016zbm,Li:2017mek,Liu:2019jka,Yu:2017beg,Walters:2017afr}. {For a recent review, see ref.~\cite{Bhandari:2021thi}.}

In addition, there is also an idea of combining the FRB data with the GW data \cite{Wei:2018cgd}. Wei et al. noticed the fact that DM is proportional to the Hubble constant $H_{0}$ while luminosity distance $d_{\rm L}$ is inversely proportional to $H_{0}$, and they found that the product of DM from FRB and $d_{\rm L}$ from GW can yield a quantity free of $H_0$, which may be useful in cosmological parameter estimation. Inspired by this work, ref.~\cite{Li:2019klc} introduced a cosmology-independent estimate of the fraction of baryon mass in the intergalactic medium (IGM). In ref.~\cite{Zhao:2020ole}, the authors (including three of the authors in the present paper) also showed that combining the FRB data with the GW data could be helpful in cosmological parameter estimation. GWs provide a new method to measure the cosmic distance \cite{Schutz:1986gp}, dubbed as ``standard sirens" \cite{Holz:2005df}, which successfully avoids the possible systematics in the cosmic distance ladder method. With
the third-generation ground-based GW detectors, such as the Einstein Telescope (ET), large amounts of GW-EM events from the mergers of binary neutron stars (BNSs) are expected to be detected \cite{Zhao:2010sz}. Thus, GWs may become a new precise cosmological probe to determine various cosmological parameters \cite{Zhang:2019loq,Zhang:2018byx,Zhang:2019ylr,Li:2019ajo,Cai:2017aea,Cai:2016sby,Aasi:2013wya,Zhao:2010sz}. Especially, the GW data could be very helpful in breaking the cosmological parameter degeneracies when combined with other cosmological probes \cite{Wang:2018lun,Wang:2019tto,Wang:2021srv,Jin:2020hmc,Jin:2021pcv,Zhao:2019gyk}.

The capability of future FRB data in improving the cosmological parameter estimation in the $w\rm{CDM}$ and Chevallier-Polarski-Linder (CPL) models was studied in ref.~\cite{Zhao:2020ole}. It has been shown that the FRB data could break the parameter degeneracies in the CMB and GW data. But these two models and the most considered models in the literature \cite{Deng:2013aga,Gao:2014iva,Wu:2020jmx,Hagstotz:2021jzu,Wu:2021jyk,Zhou:2014yta,Jaroszynski:2018vgh} are parameterized and phenomenological models, and thus it is important to consider some dark energy models with deep and solid theoretical foundations and see whether the FRB data are still useful in cosmological parameter estimation in these dark-energy theoretical models. In this work, we investigate two dark energy models, i.e., the holographic dark energy (HDE) model and the Ricci dark energy (RDE) model. The HDE model is viewed to have a quantum gravity origin, which is constructed by combining the holographic principle of quantum gravity with the effective quantum field theory \cite{Li:2004rb,Wang:2016och}. Recently, a general covariant local field theory of HDE \cite{Lin:2021bxv} and the structure formation in the effective field theory of HDE is studied \cite{Ganz:2021hmp}. The HDE model not only can naturally explain the fine-tuning and coincidence problems \cite{Li:2004rb}, but also can fit the current and historical observational data well \cite{Zhang:2005hs,Zhang:2007sh,Chang:2005ph,Li:2013dha,Xu:2016grp,Huang:2004wt,Feng:2007wn,Ma:2007pd,Akhlaghi:2018knk,Sadri:2018rcp}, and have the potential to solve some major cosmological problems, such as the Hubble tension \cite{Dai:2020rfo}. The RDE model, as a theoretical variant of HDE, uses the average radius of the Ricci scalar curvature rather than the future event horizon of the universe as the infrared (IR) cutoff within the theoretical framework of HDE \cite{Gao:2007ep,Zhang:2009un}. Although the RDE model is not favored by the current observations \cite{Xu:2016grp}, we still study it as a demonstration in the issue of forecasting the capability of FRB data in cosmological parameter estimation.

In this paper, we shall first combine the future FRB data with the current CMB data from \emph{Planck}, and then we will study the combination of two future big-data low-redshift measurements, the FRB data and the GW standard siren data, in cosmological parameter estimation in the HDE and RDE models.

This paper is organized as follows. A brief description of the HDE and RDE models, the methods for simulating the FRB data and the GW data, and the current cosmological data are introduced in Section \ref{sec:Method}. The constraint results and relevant discussions are given in Section \ref{sec:Result}. We present our conclusions in Section \ref{sec:con}.
Throughout this paper, we adopt the units in which the speed of light equals 1.

\section{Methods and data}\label{sec:Method}

\subsection{Brief description of the HDE and RDE models}

According to the Bekenstein entropy bound, an effective field theory considered in a box of size $L$ with ultraviolet (UV) cutoff ${\varLambda}_{\rm uv}$ gives the total entropy  $S=L^{3}\varLambda_{\rm uv}^{3}\le S_{\rm{BH}}\equiv \pi M^{2}_{\rm{Pl}}L^{2}$, where $S_{\rm{BH}}$ is the entropy of a black hole with the same size $L$ and $M_{\rm{Pl}}=1/\sqrt{8\pi G}$ is the reduced Planck mass. However, Cohen et al. pointed out that in quantum field theory a short distance (i.e., UV) cutoff is related to a long distance (i.e., IR) cutoff due to the limit set by forming a black hole, and proposed a more restrictive bound, i.e., the energy bound \cite{Cohen:1998zx}. If the quantum zero-point energy density $\rho_{\rm{vac}}$ is relevant to a UV cutoff, the total energy of a system with size $L$ would not exceed the mass of a black hole of the same size, namely, $L^{3}\rho_{\rm{vac}}\le LM^{2}_{\rm{Pl}}$. Obviously, the IR cutoff size of this effective quantum field theory is taken to be the largest length size compatible with this bound.

If we take the whole universe into account, the vacuum energy related to this holographic principle is viewed as dark energy, called ``holographic dark energy''. The dark energy density can be expressed as \cite{Li:2004rb}
\begin{align}
\rho_{\rm{de}}=3c^{2}M^{2}_{\rm{Pl}}L^{-2},
\end{align}
where $c$ is a dimensionless model parameter, which is used to characterize all the theoretical uncertainties in the effective quantum field theory, and this parameter is extremely important in phenomenologically determining the evolution of HDE.

If $L$ is chosen to be the Hubble scale of the universe, then the dark energy density will be close to the observational result. However, Hsu pointed out that the equation of state (EoS) of dark energy in this case is not correct \cite{Hsu:2004ri}. Li subsequently suggested that $L$ should be chosen to be the size of the future event horizon \cite{Li:2004rb},
\begin{align}
R_{\mathrm{eh}}(a)=a \int_{t}^{\infty} \frac{d t^{\prime}}{a}=a \int_{a}^{\infty} \frac{d a^{\prime}}{H(a') a^{\prime 2}},
\end{align}
where $a$ is the scale factor of the universe and $H(a)$ is the Hubble parameter as a function of $a$. In this case, the EoS of dark energy can realize the cosmic acceleration, and the model with such a setting is usually called the HDE model.

In the HDE model, the dynamical evolution of dark energy is governed by the following differential equations,
\begin{align}
\frac{1}{E(z)} \frac{d E(z)}{dz}&=-\frac{\Omega_{\mathrm{de}}(z)}{1+z}\left(\frac{1}{2}+\frac{\sqrt{\Omega_{\mathrm{de}}(z)}}{c}-\frac{3}{2 \Omega_{\mathrm{de}}(z)}\right),\\
\frac{d \Omega_{\mathrm{de}}(z)}{d z}&=-\frac{2 \Omega_{\mathrm{de}}(z)\left(1-\Omega_{\mathrm{de}}(z)\right)}{1+z}\left(\frac{1}{2}+\frac{\sqrt{\Omega_{\mathrm{de}}(z)}}{c}\right),
\end{align}
where $E(z)\equiv H(z)/H_{0}$ is the dimensionless Hubble parameter and $\Omega_{\rm{de}}(z)$ is the fractional density of dark energy. Solving these two differential equations with the initial conditions $\Omega_{\rm de}=1-\Omega_{\rm m0}$ and $E(0)=1$ will obtain the evolutions of $\Omega_{\rm{de}}(z)$ and $E(z)$.
Then from the energy conservation equations,
\begin{align}
\dot{\rho}_{\rm{m}}+3 H\rho_{\rm{m}}&=0,\n\\
\dot{\rho}_{\rm{d e}}+3H(1+w) \rho_{\rm{d e}}&=0 ,
\end{align}
where a dot represents the derivative with respect to the cosmic time $t$ and $\rho_{\rm{m}}$ is the matter density, one can get the EoS of dark energy in the HDE model,
\begin{align}
w=-\frac{1}{3}-\frac{2\sqrt{\Omega_{\rm{de}}}}{3c} .
\end{align}

The RDE model is defined by choosing the average radius of the Ricci scalar curvature as the IR cutoff length scale in the theory.
In FRW cosmology, the Ricci scalar takes
\begin{align}
R=-6\left(\dot{H}+2 H^{2}+\frac{k}{a^{2}}\right) ,
\end{align}
where $k = 1$, $0$, and $-1$ stands for closed, flat, and open geometry, respectively, and we take $k = 0$ in the rest of this work. The dark energy density in the RDE model can be expressed as \cite{Gao:2007ep,Zhang:2009un}
\begin{align}
\rho_{\mathrm{de}}=3 \gamma M_{\mathrm{pl}}^{2}\left(\dot{H}+2 H^{2}\right),
\end{align}
where $\gamma$ is a positive constant  redefined in terms of $c$. The evolution of the Hubble parameter is determined by the following differential equation,
\begin{align}
E^{2}=\Omega_{\mathrm{m}} e^{-3 x}+\gamma\left(\frac{1}{2} \frac{d E^{2}}{d x}+2 E^{2}\right),
\end{align}
with $x\equiv \ln a$. The solution to this differential equation is found to be
\begin{align}
E(z)=\left(\frac{2 \Omega_{\mathrm{m}}}{2-\gamma}(1+z)^{3}+\left(1-\frac{2 \Omega_{\mathrm{m}}}{2-\gamma}\right)(1+z)^{\left(4-\frac{2}{\gamma}\right)}\right)^{1 / 2}.
\end{align}

\subsection{Simulation of  FRBs}\label{21}

When EM waves propagate in plasma, they will interact with free electrons and generate  dispersion. The group velocities of EM waves vary with respect to frequency, resulting in lower frequency signal being delayed. By measuring the time delay $(\Delta t)$ of the pulse signal between the highest frequency $(\nu_{\rm h})$ and the lowest frequency $(\nu_{\rm l})$, we could obtain the dispersion measure of an FRB,
\begin{align}
\mathrm{D M}=\Delta t \frac{2 \pi m_{\mathrm{e}} }{e^{2}} \frac{\left(\nu_{\rm l} \nu_{\mathrm{h}}\right)^{2}}{\nu_{\mathrm{h}}^{2}-\nu_{\rm l}^{2}} ,
\end{align}
where $m_{\rm e}$ is the electron mass and $e$ is the unit charge. The physical interpretation of DM is the integral of the electron number density along the line-of-sight, which is expressed as
\begin{align}
\mathrm{D M}=\int_{0}^{D} n_{\mathrm{e}}(l) \mathrm{d} l,
\end{align}
where $n_{\mathrm{e}}$ is the electron number density, $l$ is the path length, and $D$ is the distance to FRB. The DMs of current FRB observations are mainly in the range of $100 \sim 3000$ $\mathrm{pc}~\mathrm{cm}^{-3}$ \cite{Amiri:2021tmj}, exceeding the amount of the dispersion contributed by the Milky Way by 10 to dozens of times.

Until now, the progenitors of FRBs have not been generally figured out (excluding
a Galactic magnetar), so the real redshift distribution of FRBs is still an open issue. We assume that the comoving number density distribution of FRBs is proportional to the cosmic star formation history (SFH) \cite{Hopkins:2006bw,Caleb:2015uuk}, and we thus obtain the SFH-based redshift distribution of FRBs \cite{Munoz:2016tmg},
\begin{align}
N_{\rm{SFH}}(z)=\mathcal{N}_{\rm{SFH}}\frac{\dot{\rho}_{*}{d^2_{\rm C}}(z)}{H(z)(1+z)}e^{-{d^2_{\rm{L}}}(z)/[2{d^2_{\rm{L}}}(z_{\rm cut})]},
\end{align}
where $\mathcal{N}_{\rm{SFH}}$ is a normalization factor, $d_{\rm C}$ is the comoving distance at redshift $z$, and $z_{\rm cut}=1$ is a Gaussian cutoff. The number of detected FRBs decreases at $z>z_{\rm cut}$ due to the instrumental SNR threshold effect. The parameterized density evolution $\dot{\rho}_{*}(z)$ reads
\begin{align}
\dot{\rho}_{*}(z)=l \frac{a+b z}{1+(z / n)^{d}} ,
\end{align}
with $l=0.7$, $a=0.0170$, $b=0.13$, $n=3.3$, and $d=5.3$ \citep{Hopkins:2006bw,Cole:2000ea}.

The total observed DM of an FRB consists of the contributions from the FRB's host galaxy, IGM, and the Milky Way \cite{Thornton:2013iua,Deng:2013aga},
\begin{align}\label{eq2}
\rm{DM}_{\rm{obs}}=\rm{DM}_{\rm{host}}+\rm{DM}_{\rm{IGM}}+\rm{DM}_{\rm{MW}}.
\end{align}
The second term on the right hand side,  $\rm{DM}_{\rm{IGM}}$, relates to cosmology, and its average value is expressed as
\begin{align}\label{eq3}
\langle\mathrm{DM}_{\mathrm{IGM}}\rangle=\frac{3H_0\Omega_{\rm b}f_{\mathrm{IGM}}}{8\pi G m_{\mathrm{p}}}\int_0^z\frac{\chi(z')(1+z')dz'}{E(z')},
\end{align}
with
\begin{align}
\chi(z)=Y_{\rm H}\chi_\mathrm{{e,H}}(z)+\frac{1}{2}Y_{\rm He}\chi_\mathrm{{e,He}}(z),
\end{align}
where $f_{\mathrm{IGM}}=0.83$\footnote{Figure 9 in ref.~\cite{Shull:2011aa} shows the compilation of current observational measurements of the low-redshift baryon census. Baryons in collapsed form (galaxies, groups, and clusters), circumgalactic medium, intercluster medium, and cold neutral gas (H I and He I) make up about $17\%$ of the total. Hence, $83\%$ baryons of the universe should reside in the highly diffuse ionized gas in IGM.} is the fraction of baryon mass in IGM \cite{Shull:2011aa},
$\Omega_\mathrm{b}$ is the present-day baryon fractional density, $G$ is Newton's constant, $m_{\mathrm{p}}$ is the proton mass, $Y_{\rm H}=3/4$ is the hydrogen mass fraction, $Y_{\rm He}=1/4$ is the helium mass fraction, and the terms $\chi_\mathrm{{e,H}}$ and $\chi_\mathrm{{e,He}}$ are the ionization fractions for H and He, respectively. We take $\chi_\mathrm{{e,H}}=\chi_\mathrm{{e,He}}=1$ \cite{Fan:2006dp}, since both H and He are fully ionized when $z<3$. In eqs.~(2.16) and (2.17), the electron number density in IGM depends on two factors, i.e. the baryon number density in IGM and the free electron number fraction per baryon. The former factor is expressed by the product of the baryon number density of the whole universe,  $\frac{3H_{0}^2\Omega_b}{8\pi G m_{\rm p}}$, and the proportion of baryons in IGM to the total baryons in the universe, $f_{\rm IGM}$, and the latter factor is expressed by $\chi(z)$. Thus, the value of $f_{\rm IGM}$ is unrelated to $\chi(z)$ in which both H and He fully ionized are assumed.

From eq.~(\ref{eq2}), if we could determine $\rm{DM}_{\rm{obs}}$, $\rm{DM}_{\rm{host}}$, and $\rm{DM}_{\rm{MW}}$, the last remaining term $\rm{DM}_{\rm{IGM}}$ could be measured. The total uncertainty of $\rm{DM}_{\rm{IGM}}$ is expressed as
\begin{align}\label{eq6}
\sigma_{\rm{DM}_{\rm{IGM}}}=\left[\sigma_{\rm obs}^{2}+\sigma_{\rm MW}^{2}+\sigma_{\rm IGM}^{2}
+\left(\frac{\sigma_{\rm host}}{1+z}\right)^{2} \right]^{1/2}.
\end{align}
We take the observational uncertainty $\sigma_{\rm {obs}}=1.5~{\rm {pc~cm^{-3}}}$  \cite{Petroff:2016tcr}. From the Australia Telescope National Facility pulsar catalog \cite{Manchester:2004bp}, the average $\sigma_{\rm MW}$ is about $10~{\rm {pc~cm^{-3}}}$ for the sources at high Galactic latitudes $(|b|> 10^{\circ})$. We assume that the sources at high Galactic latitudes are included in the data, since the Galactic coordinates do not affect the predicted numbers of FRBs dramatically \cite{Hashimoto:2020dud}. Due to the inhomogeneity of the baryon matter in IGM, the deviation of an individual event from the mean $\rm{DM}_{\rm{IGM}}$ is described by  $\sigma_{\rm {IGM}}$. Here we use the following formula \cite{Li:2019klc},
\begin{align}\label{sigmaigm}
\sigma_{\rm IGM}=\begin{cases}\frac{52-45z-263z^2+21z^3+582z^4}{1-4z+7z^2-7z^3+5z^4}, &z\leq 1.03,\\
 -416+270z+480z^2+23z^3-162z^4, &1.03<z\leq 1.13,\\
38\arctan [0.6z+1]+17, &z>1.13,
\end{cases}
\end{align}
which is fitted from the simulations in refs.~\citep{FaucherGiguere:2011cy,McQuinn:2013tmc}. It should be noted that some assumptions in our work may be different from those in the numerical simulation, such as the value of baryon density. However, our fiducial values of cosmological parameters are consistent with those in ref.~\cite{Li:2019klc}. The reason of neglecting the details of numerical simulation is that for a forecast study, the error is more important than the fiducial value or the expectation value of $\rm DM_{\rm{IGM}}$($z$). The constraint precision of cosmological parameters directly depends on the error of $\rm DM_{\rm{IGM}}$($z$). Hence, although we make a simplification for convenience, using the fiducial values fully consistent with the numerical simulation \cite{FaucherGiguere:2011cy,McQuinn:2013tmc} would have negligible influence on the results. $\sigma_{\rm{host}}$ is hard to estimate because it depends on the individual properties of an FRB, for example, the type of the host galaxy, the location of FRB in the host galaxy, and the near-source plasma. We take $\sigma_{\rm{host}} = 30 ~{\rm {pc~cm^{-3}}}$ as the uncertainty of ${\rm DM_{host}}$.

Now we discuss the observed event number of the FRB detection. The intrinsic event rate and luminosity distribution of FRBs \cite{Connor:2016rhf,Macquart:2018jlq,Luo:2020wfx}, the sensitivity of a receiver, and the field of view of a telescope are important for the precise prediction on a
certain experiment. However, in this work, we focus on the issue of how many localized FRBs are required to effectively constrain the cosmological parameters in the HDE and RDE models, regardless of any specific detectors. So, here we directly refer to the predictions in the literature. Recently, Lorimer pointed out that the ‘outrigger’ stations for CHIME/FRB will detect around 1000 sources over two years \cite{lorimer2021chime}.
In the future, DSA-2000 would detect and localize FRBs at a rate of $\sim 10^3 - 10^4 $ FRBs per year \cite{Hallinan:2019qyo}; HIRAX could find dozens of localized FRBs per day \cite{Newburgh:2016mwi}; ASKAP-CRAFT could localize 1500 FRBs with arcsecond accuracy over the next 5 years \cite{Bhandari:2021thi}.
All these predictions support that the assumption of $\sim10^3$ to $10^4$ localized FRBs (with redshifts) for the observation of a few years is reasonable.
Thus, we take $N_{\rm FRB}=1000$ as a normal expectation and $N_{\rm FRB}=10000$ as an optimistic expectation, with $N_{\rm FRB}$ being the event number of FRBs.


\subsection{Simulation of standard sirens}

To simulate the GW standard siren data generated by BNS mergers from ET, we first need to assume the redshift distribution of GWs \cite{Zhao:2010sz,Cai:2016sby},
\begin{align}
P(z) \propto \frac{4 \pi d_{\mathrm{C}}^{2}(z) R(z)}{H(z)(1+z)},
\end{align}
where $R(z)$ is the time evolution of the burst rate with the form \cite{Schneider:2000sg,Cutler:2009qv,Cai:2016sby}
\begin{align}
R(z)=\left\{\begin{array}{rcl}
1+2 z, & z \leq 1, \\
\frac{3}{4}(5-z), & 1<z<5, \\
0, & z \geq 5.
\end{array}\right.
\end{align}

By observing the GW waveform of a compact binary merger, one could independently determine the luminosity distance $d_{\rm L}$ to the GW source. If the redshift of the GW source is obtained through the observation of the EM counterpart, then the $d_{\rm L}$--$z$ relation can be established to study the expansion history of the universe \cite{Holz:2005df}.

An incoming GW signal $h(t)$ could be written as a linear combination of two wave polarizations in the transverse traceless gauge,
\begin{align}
h(t)=F_+(\theta, \phi, \psi)h_+(t)+F_\times(\theta, \phi, \psi)h_\times(t),\label{ht}
\end{align}
where $\psi$ is the polarization angle, ($\theta$, $\phi$) are the location angles of the GW source, and $+$ and $\times$ denote the plus and cross polarizations, respectively. The antenna pattern functions of one Michelson-type interferometer of ET are \cite{Zhao:2010sz}
\begin{align}
F_+^{(1)}(\theta, \phi, \psi)=&~~\frac{{\sqrt 3 }}{2}\Big[\frac{1}{2}(1 + {\cos ^2}\theta )\cos (2\phi )\cos (2\psi ) - \cos \theta \sin (2\phi )\sin (2\psi )\Big],\label{equa:F1}\\
F_\times^{(1)}(\theta, \phi, \psi)=&~~\frac{{\sqrt 3 }}{2}\Big[\frac{1}{2}(1 + {\cos ^2}\theta )\cos (2\phi )\sin (2\psi ) + \cos \theta \sin (2\phi )\cos (2\psi )\Big].\label{equa:F2}
\end{align}
Since the three interferometers of ET have inclined angles of $60^\circ$ with each other, the other two pattern functions are $F_{+,\times}^{(2)}(\theta, \phi, \psi)=F_{+,\times}^{(1)}(\theta, \phi+2\pi/3, \psi)$ and $F_{+,\times}^{(3)}(\theta, \phi, \psi)=F_{+,\times}^{(1)}(\theta, \phi+4\pi/3, \psi)$.



Applying the stationary phase approximation to the time domain signal $h(t)$, we can get its Fourier transform $\tilde{h}(f)$ \cite{Zhao:2010sz},
\begin{align}
\tilde{h}(f)=\mathcal{A} f^{-7 / 6} \exp \left\{i\left(2 \pi f t_{c}-\pi / 4+2 \psi(f / 2)-\varphi_{I,(2,0)}\right)\right\},
\end{align}
where ``$\sim$" above a function denotes the Fourier transform and  $\mathcal{A}$ is the amplitude in the Fourier space,
\begin{align}
\mathcal{A}=&~~\frac{1}{d_{\rm L}}\sqrt{F_+^2(1+\cos^2\iota)^2+4F_\times^2\cos^2\iota} \sqrt{5\pi/96}\pi^{-7/6}\mathcal{M}_{\rm c}^{5/6},
\label{equa:A}
\end{align}
where $\iota$ is the inclination angle between the direction of binary's orbital angular momentum and the line of sight. The observed chirp mass is defined as $\mathcal{M}_{\rm c}=(1+z)M \eta^{3/5}$, with $M=m_1+ m_2$ being the total mass of coalescing binary system with component masses $m_{1}$ and $m_{2}$, and $\eta=m_1 m_2/M^2$ being the symmetric mass ratio. $\psi(f)$ and $\varphi_{I,(2,0)}$ are given by \cite{Sathyaprakash:2009xs,Blanchet:2004bb}
\begin{align}
\psi(f)&=-\psi_{c}+\frac{3}{256 \eta} \sum_{i=0}^{7} \psi_{i}(2 \pi M f)^{(i-5) / 3},\n\\
\varphi_{I,(2,0)}&=\tan ^{-1}\left(-\frac{2 \cos (\iota) F_{\times}}{\left(1+\cos ^{2}(\iota)\right) F_{+}}\right),
\end{align}
where $\psi_{c}$ is the coalescence phase, and the detailed expressions of the Post-Newtonian (PN) coefficients $\psi_{i}$ are given in ref.~\cite{Sathyaprakash:2009xs}. Notice that here $\psi$'s are the PN coefficients, different from the one in Eqs.~(\ref{ht})--(\ref{equa:F2}), which is the polarization angle.


The total SNR of ET is
\begin{equation}
\rho=\sqrt{\sum\limits_{i=1}^{3}(\rho^{\left(i\right)})^2},
\label{euqa:sum}
\end{equation}
where $\rho_{i}=\sqrt{\left\langle \tilde{h}^{(i)},\tilde{h}^{(i)}\right\rangle}$,
with the inner product being defined as
\begin{equation}
\left\langle{a,b}\right\rangle=4\int_{f_{\rm lower}}^{f_{\rm upper}}\frac{\tilde a(f)\tilde b^\ast(f)+\tilde a^\ast(f)\tilde b(f)}{2}\frac{df}{S_{\rm n}(f)},
\label{euqa:product}
\end{equation}
$S_{\rm n}(f)$ is the one-sided noise power spectral density. We use the interpolation method to fit the sensitivity data of ET to get the fitting function of $S_{\rm n}(f)$ \cite{ETcurve-web}. We choose  $f_{\rm lower} = 1$ Hz as the lower cutoff frequency and $f_{\rm upper} = 2/(6^{3/2}2\pi M_{\rm obs})$ as the upper cutoff frequency with $M_{\rm obs}=(m_{1}+m_{2})(1+z)$ \cite{Zhao:2010sz}.


The luminosity distance $d_{\rm L}$ from GW observation is directly used to constrain cosmological parameters. The total error of $d_{\rm L}$ is expressed as
\begin{align}\label{eq11}
\sigma_{d_{\mathrm{L}}}=\sqrt{\left(\sigma_{d_{\mathrm{L}}}^{\text {inst }}\right)^{2}+\left(\sigma_{d_{\mathrm{L}}}^{\text {lens }}\right)^{2}+\left(\sigma_{d_{\mathrm{L}}}^{\mathrm{pv}}\right)^{2}},
\end{align}
where $\sigma^{\rm inst}_{d_{\rm L}}$ , $\sigma^{\rm lens}_{d_{\rm L}}$, and $\sigma^{\rm pv}_{d_{\rm L}}$ are the instrumental error, the weak-lensing error, and the peculiar velocity error of luminosity distance, respectively. In our previous work \cite{Zhao:2020ole}, we adopted an approximation to the instrumental error,  $\sigma_{d_{\rm L}}^{\rm inst}\simeq 2d_{\rm L}/\rho$, in which the factor of 2 accounts for the maximal effect of the correlation between the luminosity distance and the inclination angle \cite{li2015extracting}. Indeed, this approximation does not exactly show the covariance between the luminosity distance and other parameters and may lead to  biased results. Hence, in this work we employ a full Fisher matrix analysis to correctly account for the parameter correlations. The Fisher information matrix of the GW signal can be expressed as
\begin{align}
\boldsymbol{F}_{i j}=\left\langle\frac{\partial \boldsymbol{h}(f)}{\partial \theta_{i}}, \frac{\partial \boldsymbol{h}(f)}{\partial \theta_{j}}\right\rangle,
\end{align}
where $\boldsymbol{h}$ is given by three interferometers of ET,
\begin{align}
\boldsymbol{h}(f)=\left[\tilde{h}_{1}(f),\tilde{h}_{2}(f),\tilde{h}_{3}(f)\right]^{\mathrm{T}},
\end{align}
and $\theta_{i}$ denotes the set of nine parameters $(\mathcal{M}_{\rm c},\eta,d_{\rm L},\theta,\phi,\psi,\iota,t_{c},\psi_{c})$ for a GW event. We then could estimate $\sigma^{\rm inst}_{d_{\rm L}}$ as
\begin{align}
\sigma^{\rm inst}_{d_{\rm L}}=\Delta\theta_{d_{\rm L}}=\sqrt{\left(F^{-1}\right)_{d_{\rm L}d_{\rm L}}},
\end{align}
where $F_{ij}$ is the total Fisher information matrix for the ET with three interferometers.

The GW data used in this work are ``bright sirens" and thus need the redshift information. The redshifts of GW sources are measured by their EM counterparts. In this work, we consider the short gamma ray bursts (SGRBs) and their afterglows as EM counterparts. Since SGRBs are supposed to be strongly beamed, we limit the observations of inclination angle from SGRBs within $20^\circ$. Therefore, in the Fisher matrix for each GW event, the sky location $(\theta , \phi)$, the inclination angle $\iota$, the mass of each NS ($m_1$, $m_2$), the coalescence phase $\psi_{\rm c}$,  and the polarization angle $\psi$ are evenly sampled in the ranges of [0, $\pi$], [0, $2\pi$], [0, $\pi/9$], $[1, 2]\ M_{\odot}$, $[1, 2]\ M_{\odot}$, [0, $2\pi$], and [0, $2\pi$], respectively, where $M_{\odot}$ is the solar mass. To eliminate the scatter of mock data, these parameters are randomly selected for 100 times to perform the Fisher matrix analysis, and we adopt the average results.

We then turn to the other two errors. The weak-lensing error is given by \cite{Hirata:2010ba,Tamanini:2016zlh}
\begin{align}
\sigma_{d_{\mathrm{L}}}^{\text {lens }}(z)=d_{\mathrm{L}}(z) \times 0.066\left[\frac{1-(1+z)^{-0.25}}{0.25}\right]^{1.8}.
\end{align}
Here we introduce a delensing factor, i.e. the use of dedicated matter surveys along the line of sight of the GW event in order to estimate the lensing magnification distribution and thus remove part of the weak lensing uncertainty.
Following refs.~\cite{Speri:2020hwc,Yang:2021qge}, the delensing could be achieved 30\% at $z=2$ and the delensing factor can be expressed as
\begin{align}
F_{\text {delens }}(z)=1-\frac{0.3}{\pi / 2} \arctan \left(z / z_{*}\right),
\end{align}
with $z_{*}=0.073$. The final lensing uncertainty on $d_{\rm L}$ is
\begin{align}
\sigma_{d_{\mathrm{L}}}^{\text {wl}}(z)=F_{\text {delens }}(z) \sigma_{d_{\mathrm{L}}}^{\text {lens }}(z).
\end{align}
Here we use $\sigma_{d_{\mathrm{L}}}^{\text {wl}}$ to replace $\sigma_{d_{\mathrm{L}}}^{\text {lens }}$ in eq.~(\ref{eq11}).

The peculiar velocity of the source relative to the Hubble flow introduces an additional source of error. For the peculiar velocity error, we take the form in ref.~\cite{Kocsis:2005vv}
\begin{align}
\sigma_{d_{\mathrm{L}}}^{\mathrm{pv}}(z)=d_{\mathrm{L}}(z) \times\left[1+\frac{(1+z)^{2}}{H(z) d_{\mathrm{L}}(z)}\right] \sqrt{\left\langle v^{2}\right\rangle},
\end{align}
where $H(z)$ is the Hubble parameter and $\sqrt{\left\langle v^{2}\right\rangle}$ is the peculiar velocity of the GW source, set to be $500\, {\rm km\ s^{-1}}$.

\begin{figure*}[htbp]
\subfigure[]{\includegraphics[width=7.4cm,height=5.4cm]{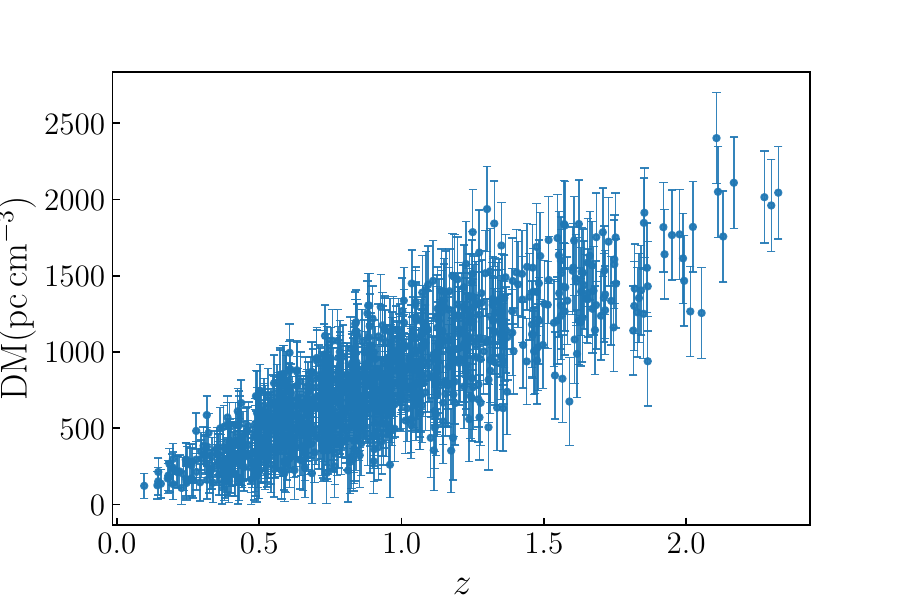}
}
\subfigure[]{\includegraphics[width=7.4cm,height=5.4cm]{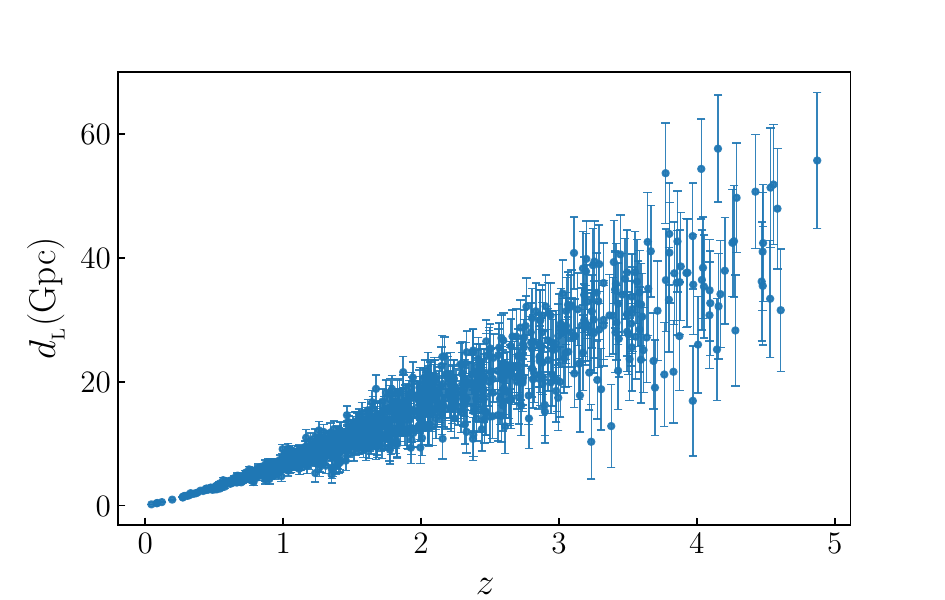}
}
\centering
 \caption{The mock FRB data (panel a) and the mock GW data from ET (panel b). The FRB data are simulated from the normal expectation scenario with 1000 FRB events, and the GW data are simulated from the 10-year observation of ET with 1000 BNS merger events. 
 }\label{FRB+ET}
\end{figure*}

GW170817 was detected by the current second-generation ground-based GW detectors and its SGRB counterpart was detected by the $Neil ~ Gehrels ~ Swift$ satellite \cite{Evans:2017mmy}. In the future, with the higher sensitivity of the third-generation ground-based GW detectors and the next-generation EM facilities, it is possible to detect much more GW events with redshift information. For example, ET is expected to detect about $10^{5}$ BNS mergers per year but in which only $\sim 10^{-3}$ of the events with $\gamma$-ray bursts are toward us \cite{Yu:2021nvx}. So, in the era of ET and CE, a few $\times 10^{2}$ GW events could be treated as ``bright sirens" per year. Recently, Chen et al. conceived a future mission ``$Swift++$'' and forecasted that about 910 GW bright siren events could be detected by CE and $Swift++$ with a 10-year observation \cite{Chen:2020zoq}.
In this work, we simulate 1000 standard siren events generated by BNS mergers within a 10-year observation of ET.  In figure \ref{FRB+ET}, we make a comparison between the FRB mock data in the normal expectation scenario and the GW mock data. We can see that in our simulations, the FRB data are mainly distributed below $z\approx 1.5$, but the GW data could extend to $z\approx 3$.

\subsection{Cosmological data}

For the current mainstream cosmological data, we use the \emph{Planck} CMB ``distance priors" derived from the \emph{Planck} 2018 data release \cite{Chen:2018dbv}, and the BAO measurements from 6dFGS at $z_{\rm eff} = 0.106$ \cite{Beutler:2011hx}, SDSS-MGS at $z_{\rm eff} = 0.15$ \cite{Ross:2014qpa}, and BOSS-DR12 at $z_{\rm eff} = 0.38$, 0.51, and 0.61 \cite{Alam:2016hwk}. In the generation of the FRB and GW mock data, the fiducial values of cosmological parameters are taken to be the best-fit values of CMB+BAO+SN from ref.~\cite{Zhang:2019ple}. We use the Markov-chain Monte Carlo analysis \cite{Lewis:2002ah} to obtain the posterior probability distribution of the cosmological parameters.

\section{Results and discussion} \label{sec:Result}

\subsection{CMB+FRB}\label{CF}

\begin{table*}[htbp]
\setlength\tabcolsep{5.0pt}
\renewcommand{\arraystretch}{1.5}
\centering
\begin{tabular}{cccccccc}
\hline
          Model          & Error   & CMB & CMB+BAO & FRB1 & CMB+FRB1& FRB2& CMB+FRB2 \\ \hline
\multirow{4}{*}{HDE} & $\sigma(\Omega_{\rm m})$ &
$0.032$ & $0.012$ & $0.062$ & $0.022$ & $0.026$& $0.012$\\
& $\sigma(h)$ &
$0.035$ & $0.013$ & $-$ & $0.022$&$-$&$0.012$\\
& $\sigma(c)$ &
$0.18$ & $0.087$ & $0.55$ & $0.11$&$0.44$&$0.057$\\
& $10^2\sigma(\Omega_{\rm b} h^2)$ &
$0.015$ & $0.015$ & $0.56$ & $0.014$&$0.53$&$0.0097$\\ \hline
\multirow{4}{*}{RDE} & $\sigma(\Omega_{\rm m})$ &
$0.060$ & $-$ & $0.087$ & $0.043$&$0.032$&$0.020$\\
& $\sigma(h)$ &
$0.057$ & $-$ & $-$ & $0.045$&$-$&$0.021$\\
& $\sigma(\gamma)$ &
$0.019$ & $-$ & $0.24$ & $0.014$&$0.058$&$0.0081$\\
& $10^2\sigma(\Omega_{\rm b} h^2)$ &
$0.015$ & $-$ & $0.67$ & $0.013$&$0.51$&$0.0076$\\ \hline
\end{tabular}
\caption{Absolute errors ($1\sigma$) of the cosmological parameters in the HDE and RDE models by using the CMB, CMB+BAO, FRB1, CMB+FRB1, FRB2, and CMB+FRB2 data. FRB1 and FRB2 denote the FRB data in the normal expectation scenario (i.e., $N_{\rm FRB}=1000$) and optimistic expectation scenario (i.e., $N_{\rm FRB}=10000$), respectively.}
\label{tab:full}
\end{table*}

\begin{figure*}[htbp]
\subfigure[]{\includegraphics[width=7.2cm,height=5.4cm]{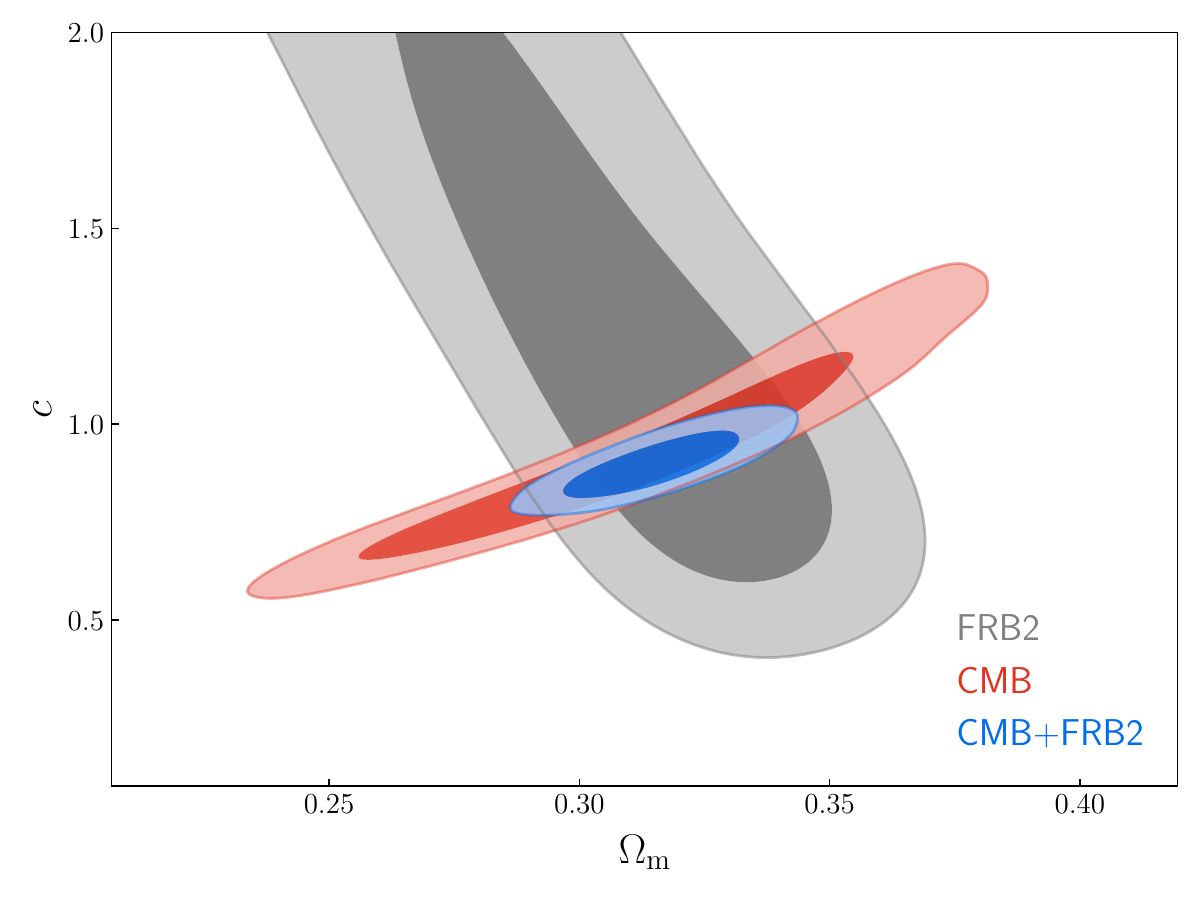}}
\subfigure[]{\includegraphics[width=7.2cm,height=5.4cm]{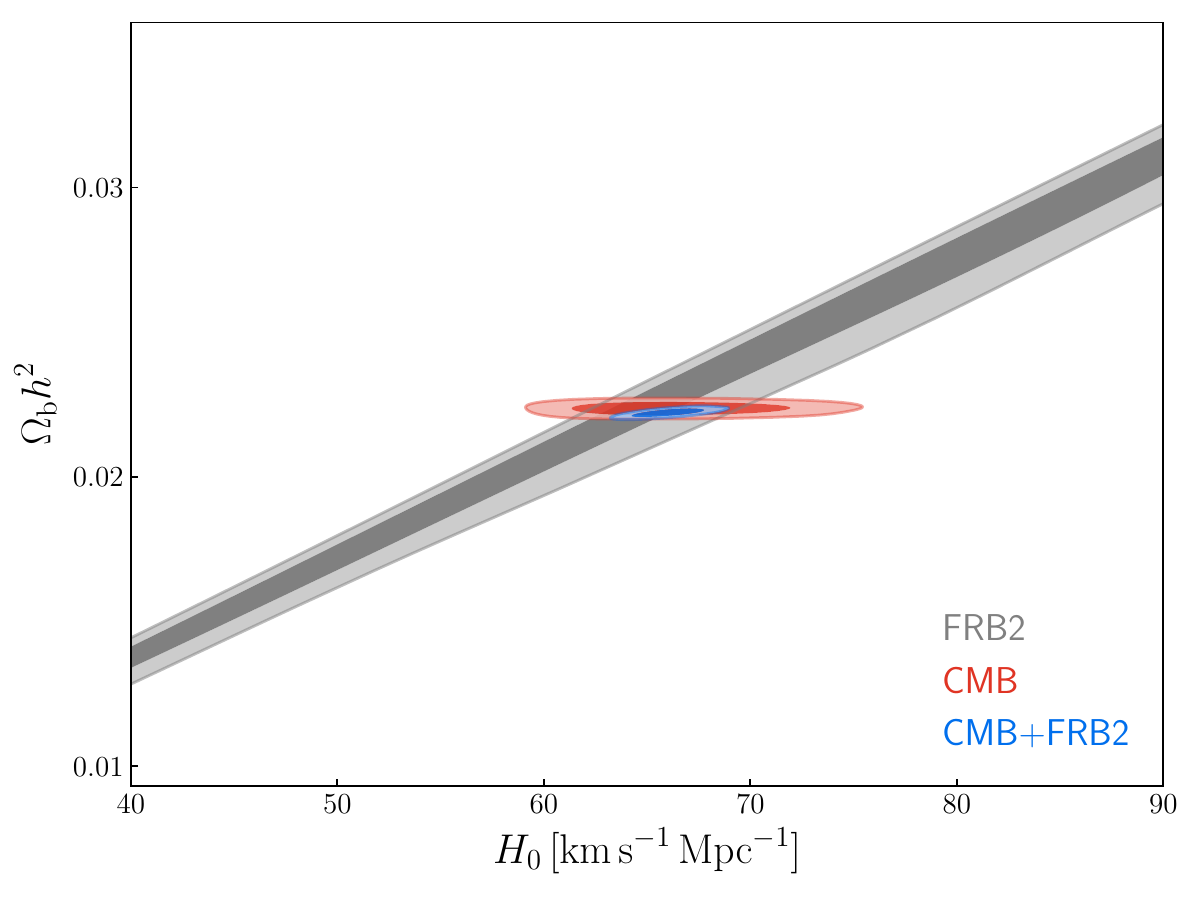}}
\centering
 \caption{\label{HDE}Two-dimensional marginalized contours (68.3\% and 95.4\% confidence levels) in the $\Omega_{\rm m}$--$c$ plane (panel a) and the $H_{0}$--$\Omega_{\rm b}h^2$ plane (panel b) for the HDE model, by using the FRB, CMB, and CMB+FRB data. Here, the FRB data are simulated based on the optimistic expectation scenario.}
\end{figure*}

\begin{figure*}[htbp]
\subfigure[]{\includegraphics[width=7.2cm,height=5.4cm]{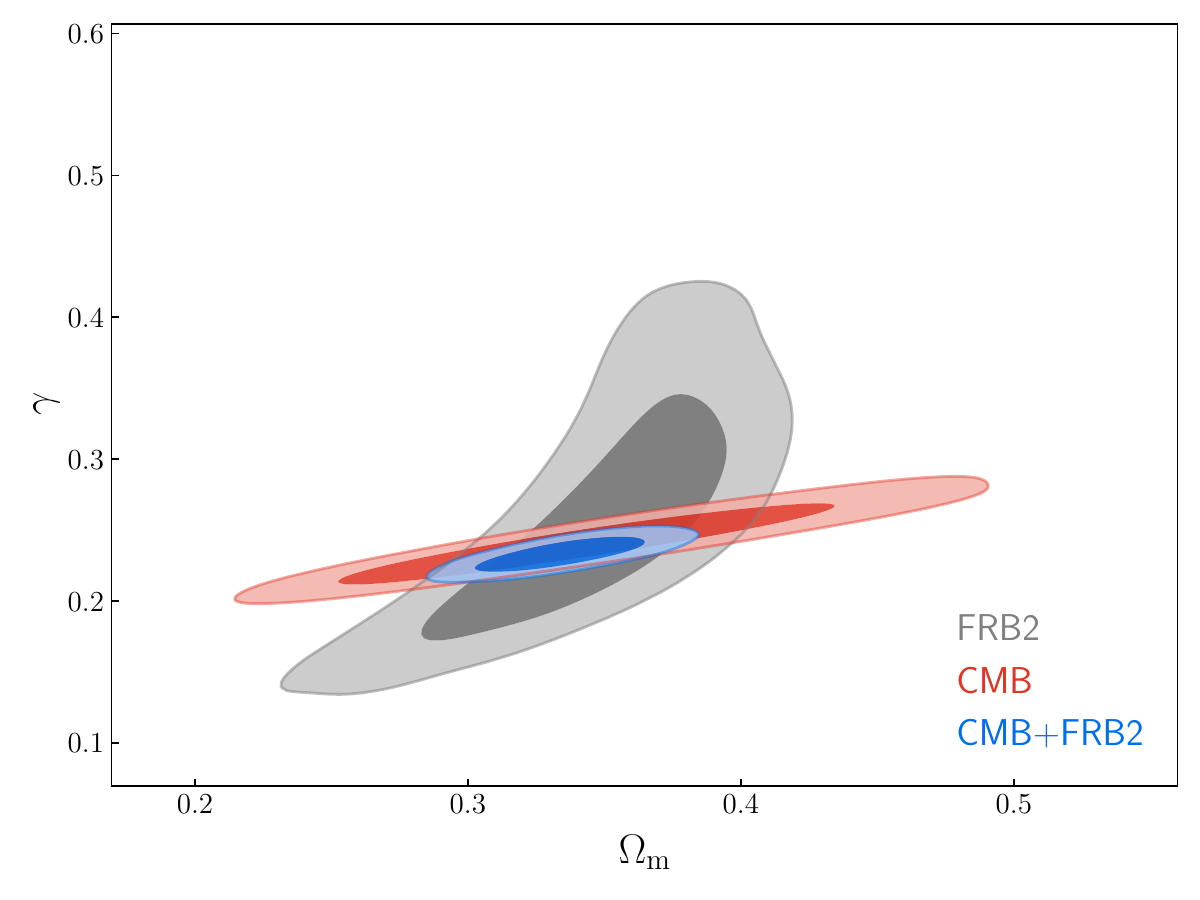}}
\subfigure[]{\includegraphics[width=7.2cm,height=5.4cm]{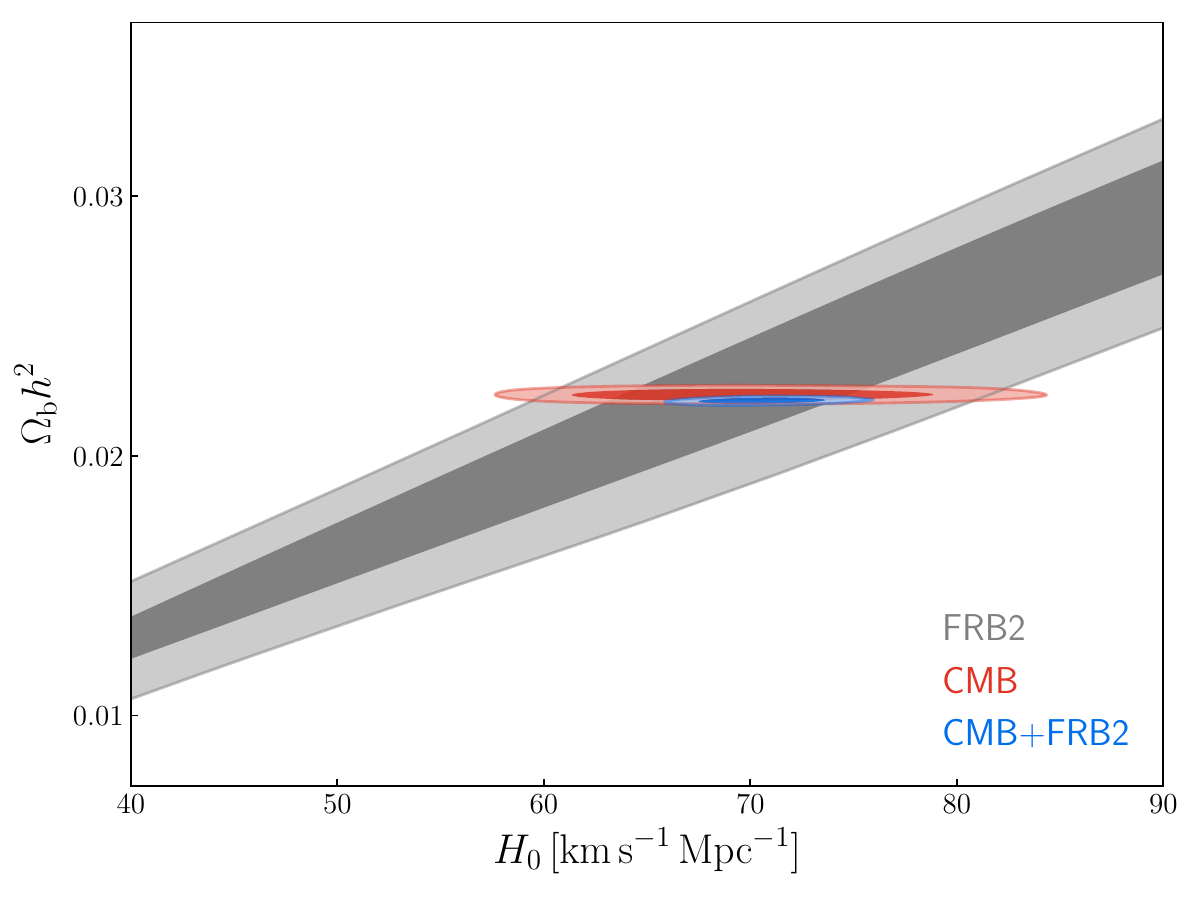}}
\centering
 \caption{ \label{RDE}Two-dimensional marginalized contours (68.3\% and 95.4\% confidence levels) in the $\Omega_{\rm m}$--$\gamma$ plane (panel a) and the $H_{0}$--$\Omega_{\rm b}h^2$ plane (panel b) for the RDE model, by using the FRB, CMB, and CMB+FRB data. Here, the FRB data are simulated based on the optimistic expectation scenario.}
\end{figure*}

In this subsection, we investigate the capability of the FRB data of breaking the parameter degeneracies inherent in the CMB data. We shall first discuss the improvement of the precision of cosmological parameters with the addition of the FRB data. Then we compare the capabilities of FRB and of BAO data in breaking the parameter degeneracies. The absolute errors (1$\sigma$) of the cosmological parameters in the HDE and RDE models are listed in table~\ref{tab:full}. We use FRB1 and FRB2 to denote the FRB data in the normal expectation (i.e., $N_{\rm FRB}=1000$) and the optimistic expectation (i.e., $N_{\rm FRB}=10000$), respectively.
Here, for a cosmological parameter $\xi$, we use $\sigma(\xi)$ to denotes its absolute error, and we also use the relative error $\varepsilon(\xi)=\sigma(\xi)/\xi$ in the following discussions.

From table~\ref{tab:full} we see that the cosmological constraints from the FRB1 and FRB2 data are looser than those from the CMB data (except for $\Omega_{\rm m}$ from FRB2). However, with combining the CMB and FRB data, the constraints from the CMB+FRB1 and CMB+FRB2 data are both evidently improved compared with those from the CMB data alone. As shown in figure~\ref{HDE}, in the HDE model, $\Omega_{\rm m}$ and $c$ are positively correlated in the CMB data, but they are anti-correlated in the FRB data. Combining the CMB and FRB data breaks the parameter degeneracies, and thus improves the constraints on the EoS parameters of dark energy and the matter density parameter. For the HDE model, the CMB data combined with the FRB1 and FRB2 data give the results $\varepsilon(c)=11.7\%$ and $\varepsilon(c)=6.3\%$, respectively. The absolute errors of $c$ are reduced by about 40.0\% and 67.7\% by combining the FRB1 and FRB2 data with the CMB data, respectively. It is worth emphasizing that the constraint  $\varepsilon(c)=6.3\%$ is close to the constraint precision given by the CMB+BAO+SN data \cite{Zhang:2019ple}. That is to say, using only the future FRB data combined with the CMB data could provide precise cosmological constraints comparable with the current mainstream data.
For the RDE model, it is shown in figure~\ref{RDE} that the FRB data also break the parameter degeneracies inherent in the CMB data. The CMB data combined with the FRB1 and FRB2 data give $\varepsilon(\gamma)=5.9\%$ and $\varepsilon(\gamma)=3.5\%$, respectively. The absolute errors of $\gamma$ are reduced by about 24.3\% and 56.2\%, by adding the FRB1 and FRB2 data to the CMB data, respectively.

Then we turn our attention to the constraints on the baryon density $\Omega_{\rm b}$ and the Hubble constant $H_0$. Using big bang nucleosynthesis (BBN) and CMB can precisely determine the value of  $\Omega_{\rm b}h^2$, however, in the nearby universe, the observed baryons in stars, the cold interstellar medium, residual Ly$\alpha$ forest gas, \textsc{O\,vi}, broad \textsc{H\,i} Ly$\alpha$ absorbers, and hot gas in clusters of galaxies account for only $\sim 50\%$ of the baryons \cite{Bhandari:2021thi,driver2021challenge}. This is called the missing baryon problem.
On the other hand, for the measurements of the Hubble constant, in recent years there appeared the puzzle of ``Hubble tension''.
The values of the Hubble constant derived from different observations show strong tension, which actually reflects the inconsistency of measurements between the early universe and the late universe \cite{Riess:2020sih,Verde:2019ivm,Guo:2018ans,cai:2020,Liu:2019awo,Zhang:2019cww,Ding:2019mmw,Guo:2019dui,Guo:2018uic,Feng:2019jqa,Gao:2021xnk}.
One may expect to precisely measure $\Omega_{\rm b}h^2$ and $H_0$ with powerful low-redshift probes.


In figure~\ref{HDE}(b), we show the marginalized posterior probability distribution contours in the $H_0$--$\Omega_{\rm b}h^2$ plane for the HDE model. We see that $H_0$ and $\Omega_{\rm b}h^2$ are strong positively correlated when using the FRB data alone, because $\mathrm{DM}_{\mathrm{IGM}}$ is proportional to $H_0 \Omega_{\rm b}$ [see eq.~(\ref{eq3})]. It is found that solely using the FRB2 data cannot constrain $H_0$, but can effectively constrain $\Omega_{\rm b}h^2$, giving the result of $\sigma(\Omega_{\rm b}h^2)\approx 0.005$, which is an about 22.7\% constraint.
Although the precision is still low, the method of using the FRB observation has been proven to provide a potential way to solve the missing baryon problem.
The CMB data can place tight constraint on $\Omega_{\rm b}h^2$, and thus the data  combination FRB+CMB can give a precise measurement on $H_0$. For the HDE model, the simulated FRB1 data combined with the current CMB data could give the result $\varepsilon(h)=3.3\%$, which is reduced by 37.1\% compared to using the CMB data alone. The data combination CMB+FRB2 can give the result $\varepsilon(h)=1.8\%$, which roughly equals the measurement precision of the SH0ES value \cite{Riess:2019cxk}. For CMB+FRB2 in the RDE model, the FRB2 data combined with the CMB data could give $\varepsilon(h)=2.9\%$, and the error of $h$ are reduced by about 64\% by adding the FRB2 data to the CMB data.

\begin{figure*}[htbp]
\includegraphics[width=8cm,height=6cm]{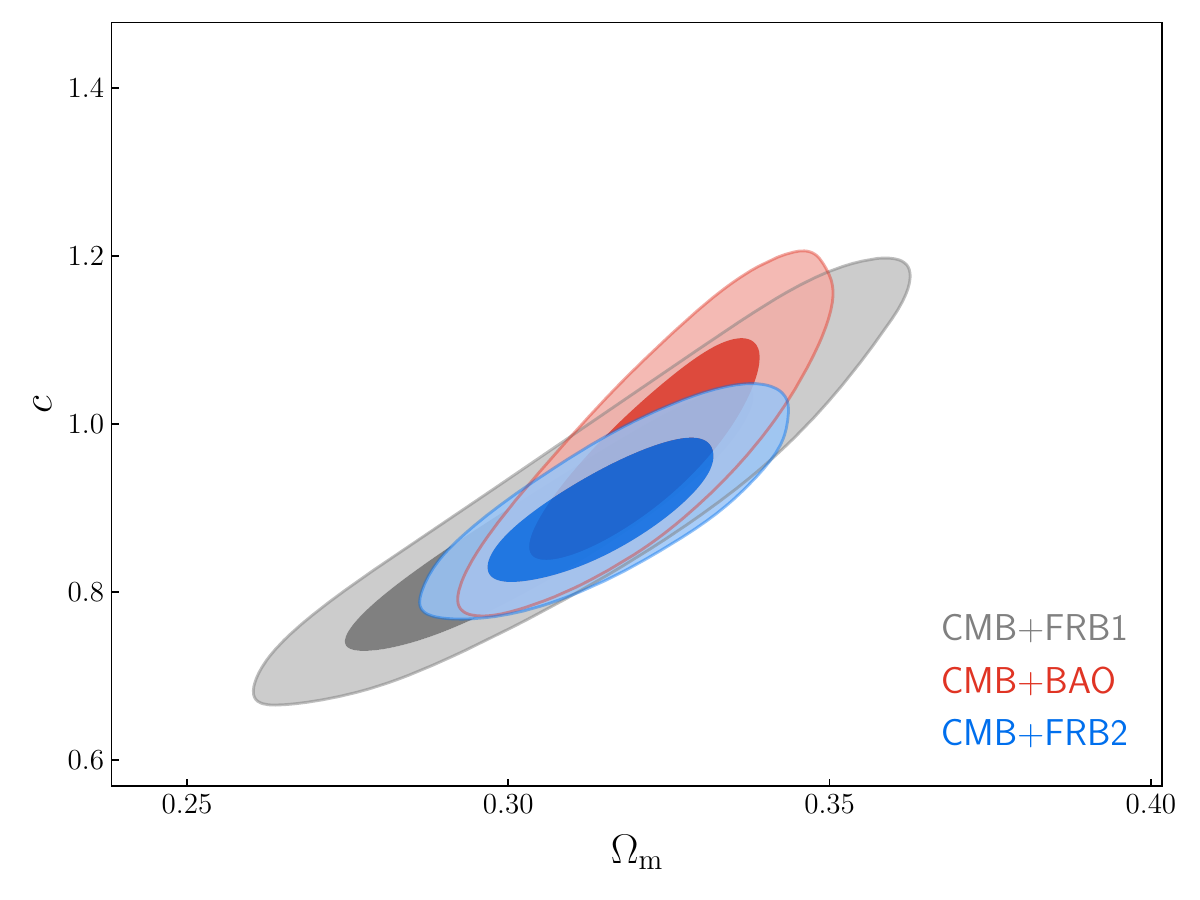}
\centering
{\caption{ \label{BAO}Two-dimensional marginalized contours (68.3\% and 95.4\% confidence levels) in the $\Omega_{\rm m}$--$c$ plane for the HDE model, by using the CMB+FRB1, CMB+BAO, and CMB+FRB2 data.}}
\end{figure*}

Moreover, we also compare the capabilities of the CMB+FRB and CMB+BAO data of constraining cosmological parameters. From figure~\ref{BAO} we can see that the constraints from the CMB+FRB2 data are comparable to those from the CMB+BAO data, even slightly better. In the HDE model, both the CMB+FRB2 and CMB+BAO data give a constraint of $\sigma(\Omega_{\rm m})=0.012$.
For the parameter $c$, the CMB+FRB2 and CMB+BAO data give the results $\sigma(c)=0.056$ and $\sigma(c)=0.087$, respectively. It indicates that the FRB2 data have  similar capability of breaking parameter degeneracies as the BAO data. 

\begin{figure*}[htbp]
\subfigure[]{\includegraphics[width=7.2cm,height=5.4cm]{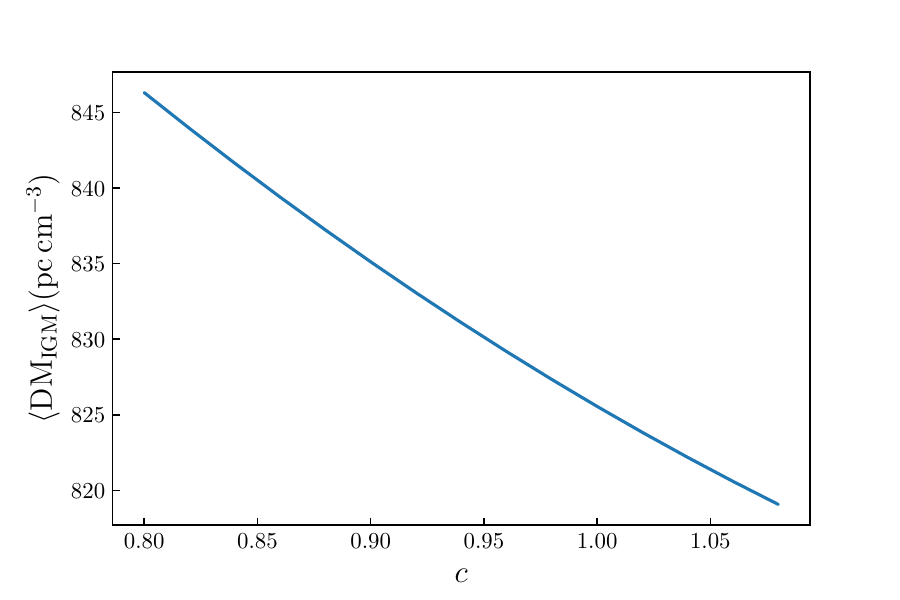}
}
\subfigure[]{\includegraphics[width=7.2cm,height=5.4cm]{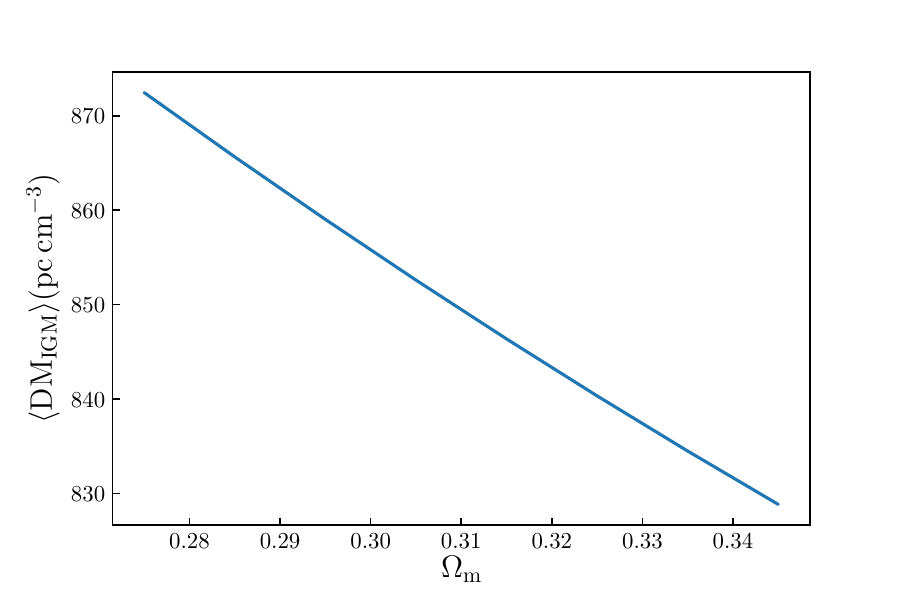}
}
\centering
 \caption{$\langle\mathrm{DM}_{\mathrm{IGM}}\rangle$ as a function of $c$ (panel a) and $\Omega_{\rm m}$ (panel b) calculated at $z = 1$. Other cosmological parameters are fixed to the fiducial values.}
 \label{DMcde}
\end{figure*}

Finally, we explain why combining the CMB and FRB data can break the parameter degeneracies. The correlation between parameters $c$ and $\Omega_{\rm m}$ is related to the used observational data. Different data measure different physical quantities, thus leading to different correlations between parameters. For the FRB data, we can plot $\langle\mathrm{DM}_{\mathrm{IGM}}\rangle$ as a function of $c$ ($\Omega_{\rm m}$) while all the other parameters are fixed, as shown in the left (right) panel of Fig.~\ref{DMcde}. The results are calculated at the mean value of our mock FRB data's redshift, $z = 1$, as a representative. We find that the parameters $c$ and $\Omega_{\rm m}$ are both inversely proportional to $\langle\mathrm{DM}_{\mathrm{IGM}}\rangle$. Therefore, $c$ and $\Omega_{m}$ are anti-correlated when using the FRB data to constrain them, since $\langle\mathrm{DM}_{\mathrm{IGM}}\rangle$ is the effective observational quantity and should be approximately viewed as a fixed value at a specific redshift.

The correlation between parameters $c$ and $\Omega_{\rm m}$ in the CMB data can also be evaluated in the above method. Here we directly refer to Fig. 3 in ref.~\cite{Li:2013dha} to show that $c$ and $\Omega_{\rm m}$ are positively correlated when they are constrained by CMB data.  The different correlations between $c$ and $\Omega_{\rm m}$ in the FRB and CMB data thus lead to the parameter degeneracies being broken.

\subsection{GW+FRB}

\begin{table*}[!htb]
\setlength\tabcolsep{9.0pt}
\renewcommand{\arraystretch}{1.5}
\centering
\begin{tabular}{cccccc}
\hline
                & Error   & GW & GW+FRB  &  CMB+GW& CMB+GW+FRB \\ \hline
\multirow{4}{*}{HDE} & $\sigma(\Omega_{\rm m})$ &
0.024 & 0.016  & 0.0059& 0.0052\\ 
& $\sigma(h)$ & 0.010& 0.0095  & 0.0064& 0.0054\\ 
& $\sigma(c)$ & 0.21 & 0.16  & 0.045 & 0.035 \\ 
& $10^2\sigma(\Omega_{\rm b} h^2)$ & $-$ & 0.011 & 0.015 &0.0074\\
\hline
\multirow{4}{*}{RDE} & $\sigma(\Omega_{\rm m})$ &
0.012 & 0.011  & 0.0090& 0.0082\\ 
& $\sigma(h)$ & 0.015& 0.014  & 0.0093& 0.0086\\ 
& $\sigma(\gamma)$ & 0.028 & 0.020  & 0.0067 & 0.0059 \\ 
& $10^2\sigma(\Omega_{\rm b} h^2)$ & $-$ & 0.011 & 0.014 &0.0069\\
\hline
\end{tabular}
\caption{Absolute errors ($1\sigma$) of the cosmological parameters in the HDE and RDE models by using the GW, GW+FRB, CMB+GW, and CMB+GW+FRB data. Here, the FRB data are simulated based on the optimistic expectation scenario.}
\label{tab:GW}
\end{table*}

\begin{figure*}[htbp]
\subfigure[]{\includegraphics[width=7.2cm,height=5.4cm]{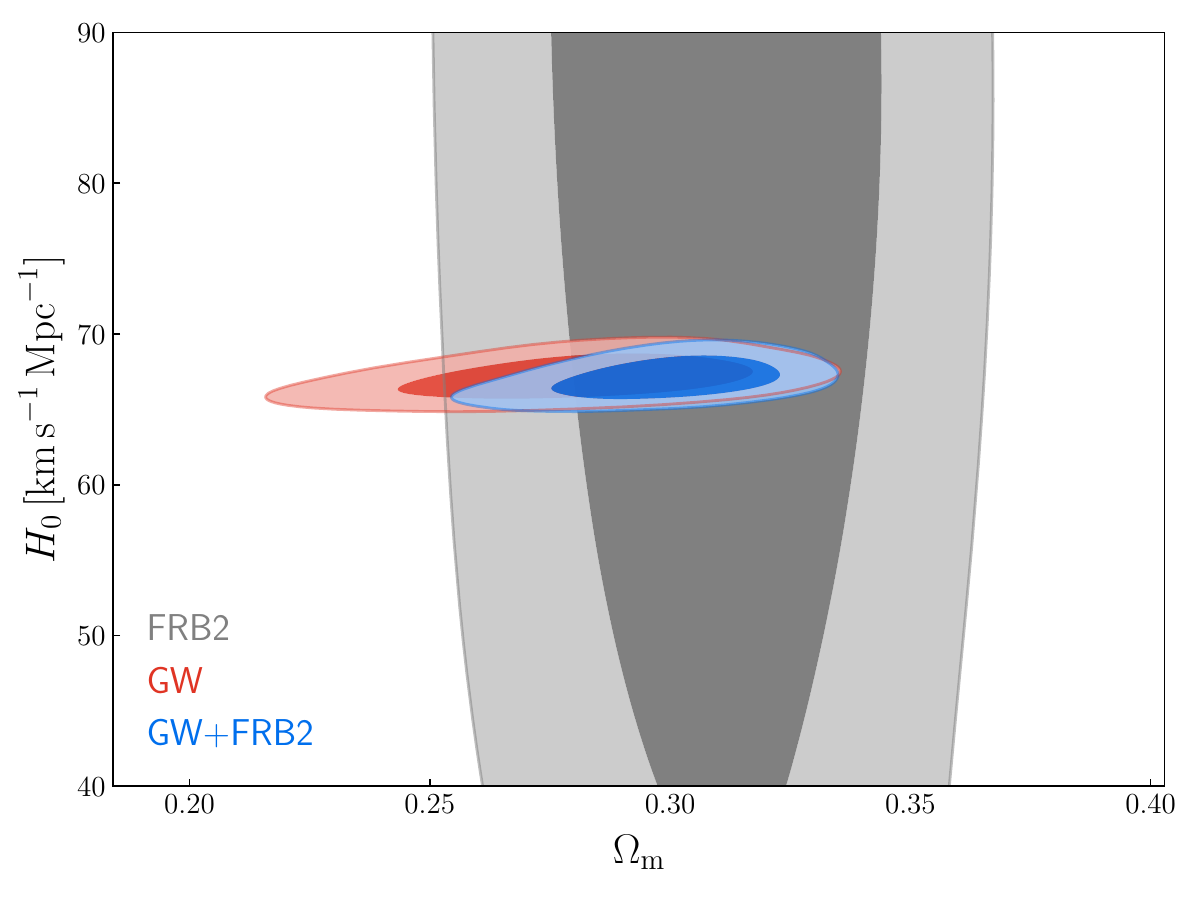}}
\subfigure[]{\includegraphics[width=7.2cm,height=5.4cm]{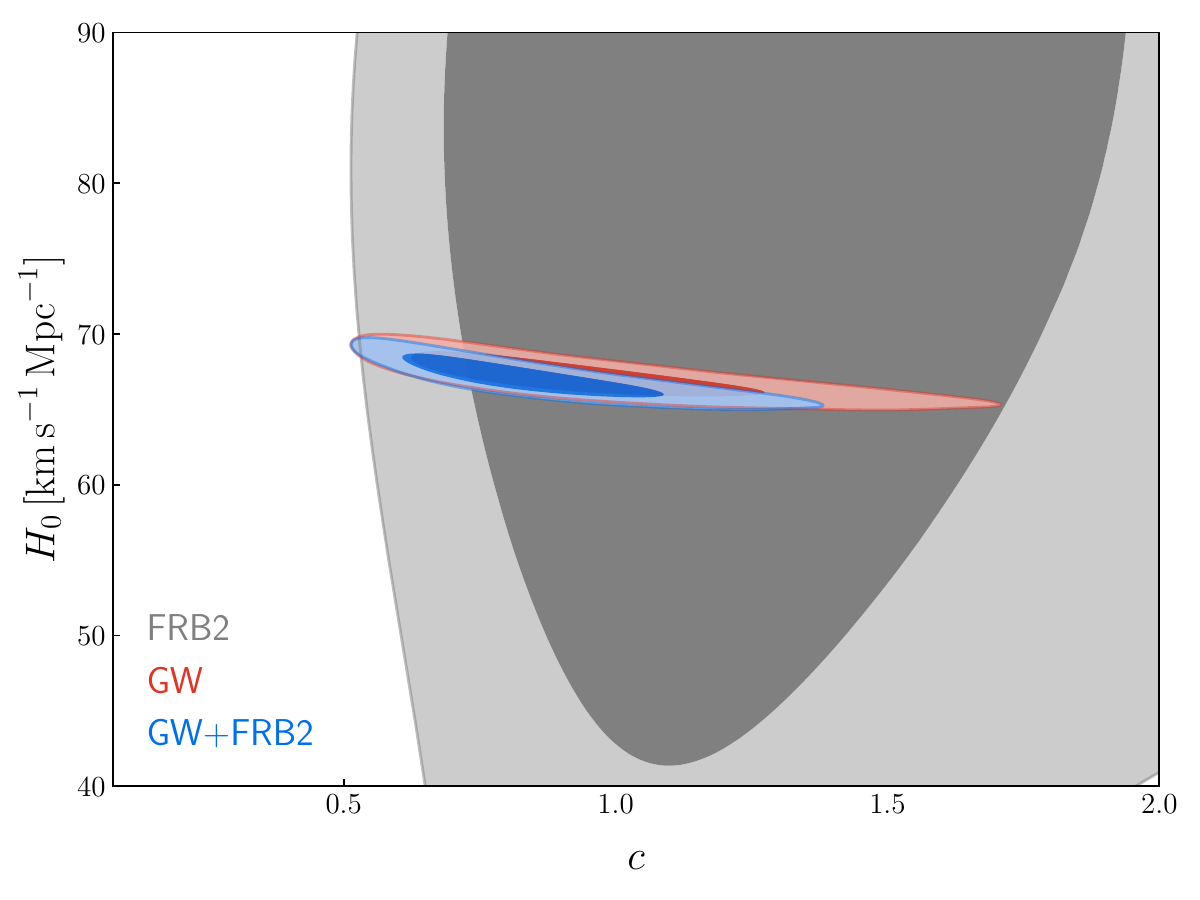}}
\centering
 \caption{ \label{GWFRB}Two-dimensional marginalized contours (68.3\% and 95.4\% confidence levels) in the $\Omega_{\rm m}$--$H_{0}$ plane (panel a) and the $c$--$H_{0}$ plane (panel b) for the HDE model, by using the FRB, GW, and GW+FRB data. Here, the FRB data are simulated based on the optimistic expectation scenario.}
\end{figure*}


In addition to the FRB observation, the GW observation is another powerful cosmological probe. As demonstrated in ref.~\cite{Zhao:2020ole}, the FRB and GW observation could help each other to break the parameter degeneracies in the $w$CDM and CPL models. In this subsection, we further study these two cosmological probes in the HDE and RDE models, and the FRB data are simulated based on the optimistic expectation scenario.

From table~\ref{tab:GW}, we see that the constraints from the data combination GW+FRB are slightly improved compared with the GW data alone. But from figure~\ref{GWFRB} we find that the parameter degeneracy orientations formed by GW and by FRB are nearly orthogonal. The inclusion of the FRB data reduces the relative error on $\Omega_{\rm m}$ from $8.4\%$ to $5.2\%$ in the HDE model. The constraints on $\Omega_{\rm m}$ and $\Omega_{\rm b}h^2$ from the GW+FRB data are comparable to those from the CMB+BAO data. However, neither the GW data nor the FRB data can give tight constraints on the parameter $c$. The data combination GW+FRB gives  $\varepsilon(c)=18.0\%$, which is only slightly better than the constraint with CMB alone. For the Hubble constant $H_0$, the FRB data actually contribute a little to the joint constraint of GW+FRB, since the GW data alone are able to precisely constrain $H_0$ but FRB alone cannot constrain it. Compared with the results in the  $w\rm{CDM}$ model \cite{Zhao:2020ole}, the improvement of parameter constraint by adding 10000 FRB data into the GW data is weaker in the HDE model. For example, the improvements of the constraints on $\Omega_{\rm m}$ are $53\%$ and $33\%$ in the $w\rm{CDM}$ and HDE models, respectively.
Therefore, we find that, for constraining the HDE model, much more FRB data are needed to be combined with the GW data.


The results from the data combination CMB+GW+FRB further confirm this statement. The inclusion of the FRB data in the data combination CMB+GW+FRB tinily improves the constraints on $\Omega_{\rm m}$ and $H_0$, and the contributions to the constraints mainly come from CMB+GW. Nevertheless, the CMB+GW data can only constrain $c$ to the precision of around 5\%, which is still far away from the standard of precision cosmology. Compared to the GW data, the FRB data are more likely to be enlarged to a larger sample. Furthermore, the event rate of FRBs detected by SKA could be at least 2--3 orders of magnitude larger than the sample size we have considered \cite{Hashimoto:2020dud}. With the accumulation of more abundant and precise data, FRBs would provide a tight constraint on the EoS of dark energy, so
we expect that in the future the large amounts of FRBs observed by SKA may have the potential to help the joint constraint on $c$ achieve the precision around 1\%.


\subsection{Effect of $\rm DM_{host}$ uncertainty}
In the above analyses, we have assumed that the $\rm DM_{host}$ uncertainty is $30\, {\rm pc\,cm^{-3}}$. The progenitors of FRBs are actually not clear and some factors are still open issues. The treatment of $\sigma_{\rm host} = 30\, {\rm pc\,cm^{-3}}$ thus could be regarded as an optimistic scenario. In this subsection, we perform an analysis based on a more conservative scenario with $\sigma_{\rm host} = 150\, {\rm pc\,cm^{-3}}$. The FRB data of this conservative scenario is represented by FRB3. The constraint results of cosmological parameters are shown in table~\ref{tab:full2}.



\begin{table*}[!htb]
\setlength\tabcolsep{8.0pt}
\renewcommand{\arraystretch}{1.5}
\centering
\begin{tabular}{cccccccc}
\hline
       Model        & Error  & FRB3 & CMB+FRB3  \\ \hline
\multirow{4}{*}{HDE} & $\sigma(\Omega_{\rm m})$ &
$0.025$ & $0.013$  \\
& $\sigma(h)$ &
$-$ & $0.013$ \\
& $\sigma(c)$ &
$0.44$ & $0.058$ \\
& $10^2\sigma(\Omega_{\rm b} h^2)$ &
$0.51$ & $0.010$ \\ \hline
\multirow{4}{*}{RDE} & $\sigma(\Omega_{\rm m})$ &
$0.040$ & $0.024$ \\
& $\sigma(h)$ &
$-$ & $0.023$ \\
& $\sigma(\gamma)$ &
$0.076$ & $0.0089$ \\
& $10^2\sigma(\Omega_{\rm b} h^2)$ &
$0.54$ & $0.0080$ \\ \hline
\end{tabular}
\caption{Absolute errors ($1\sigma$) of the cosmological parameters in the HDE and RDE models by using the FRB3 and CMB+FRB3 data. Here, FRB3 denotes the FRB data with the ${\rm DM_{host}}$ uncertainty $\sigma_{\rm host} = 150\, {\rm pc\,cm^{-3}}$.}
\label{tab:full2}
\end{table*}

From table~\ref{tab:full2}, we see that the constraints from FRB3 are slightly looser than those from FRB2. Taking the HDE model as an example, we can see that FRB3 and CMB+FRB3 give the results $\varepsilon(c)=37.3\%$ and $\varepsilon(c)=6.7\%$, respectively, which are 2.8\% and 6.0\% larger than those from FRB2 and CMB+FRB2, respectively. The influence on the constraints in the RDE model is slightly larger. The relative errors in the cases of FRB3 and CMB+FRB3 are $\varepsilon(\gamma)=27.9\%$ and $\varepsilon(\gamma)=3.8\%$, which are 20.4\% and 8.7\% larger than those from FRB2 and CMB+FRB2, respectively. From eq.~(\ref{eq6}), we see that $\sigma_{\rm host}$ contributes to the total uncertainty of $\rm{DM}_{\rm{IGM}}$ with a factor $1/(1+z)$, thus even though $\sigma_{\rm host}$ varies a lot, its impact on FRBs' capability of constraining cosmological parameters and breaking parameter degeneracies is slight. So, there may be some other factors affecting the constraints, but the main conclusions will still hold. This confirms the discussion in ref.~\cite{Zhao:2020ole}.



\section{Conclusion} \label{sec:con}
In this paper, we investigate the capability of future FRB data in improving the cosmological parameter estimation and  how many FRB data are required to effectively constrain the cosmological parameters in the HDE and RDE models. We consider two FRB scenarios: the normal expectation scenario with 1000 localized FRB data and the optimistic expectation scenario with 10000 localized FRB data.

We find that, in the HDE model, combining the FRB data and the CMB data could break the parameter degeneracies, and the joint constraints on $H_0$ and dark-energy parameters are quite tight.
To achieve the constraint precision of $H_0$ comparable with the SH0ES result \cite{Riess:2019cxk}, around 10000 FRB data are need to be combined with the CMB data. For the EoS of dark energy, 10000 FRB data combined with the CMB data would give a $\sim 6\%$ constraint, which is close to the precision given by the CMB+BAO+SN data. The data combination FRB+CMB also give a tight constraint on $\Omega_{\rm b}h^2$. The results in the RDE model also support the conclusion above. These results confirm the conclusion in ref.~\cite{Zhao:2020ole} that the future $10^4$ FRB data can become useful in cosmological parameter estimation when combined with CMB.


We also consider another powerful low-redshift cosmological probe, the GW standard sirens observation, and investigate the capability of constraining cosmological parameters when combining the FRB data with the GW data, which is independent of CMB. We use a full Fisher matrix analysis to account for the parameter correlations and avoid biased results. The inclusion of the FRB data in the data combination GW+FRB only improves the constraints slightly, but we show that the orientations of the parameter degeneracies formed by FRB and by GW are nearly orthogonal. We show that in some dark energy models such as the HDE model, 10000 FRB events are possibly not enough to be used to constrain cosmological parameters when combined with the GW data. However, we can expect that, in the near future, large amounts of localized FRBs will be observed by SKA. If we could observe much more FRBs with higher precision in the future, the combination with the GW data could significantly improve the constraints. It can be expected that FRBs observed by SKA may have potential to help the joint constraint on $c$ achieve the precision around 1\%, reaching the standard of precision cosmology.



Finally, we investigate the effect of the ${\rm DM_{host}}$ uncertainty on the constraint results. A larger ${\rm DM_{host}}$ uncertainty surely looses the constraints, but the main conclusions still hold.

\acknowledgments
We thank Zheng-Xiang Li, He Gao, Jing-Zhao Qi, Shang-Jie Jin, Li-Yang Gao, Peng-Ju Wu, Hai-Li Li, Yi-Chao Li, and Di Li for helpful discussions. This work was supported by the National Natural Science Foundation of China (Grants Nos. 11975072, 11835009, 11875102, and 11690021), the Liaoning Revitalization Talents Program (Grant No. XLYC1905011), the Fundamental Research Funds for the Central Universities (Grant No. N2005030), the National Program for Support of Top-Notch Young Professionals (Grant No. W02070050), the Science Research Grants from the China Manned Space Project (Grant No. CMS-CSST-2021-B01), and the National 111 Project of China (Grant No. B16009).

\bibliography{frbhde}{}
\bibliographystyle{JHEP}

\end{document}